\documentclass[sigconf,screen]{acmart} 
\AtBeginDocument{%
  \providecommand\BibTeX{{%
    \normalfont B\kern-0.5em{\scshape i\kern-0.25em b}\kern-0.8em\TeX}}}

\usepackage{enumitem}

\usepackage{xcolor}
\usepackage{multirow}
\usepackage{array}
\usepackage{siunitx}
\usepackage{graphicx}

\usepackage{dblfloatfix}

\usepackage{footmisc} 

\renewcommand{\paragraph}[1]{\vspace{0.3em}\noindent\textbf{\textit{#1}}\hspace*{.3em}}
\renewenvironment{quote}
  {\list{}{\rightmargin=0.5cm \leftmargin=0.5cm}%
   \item\relax}
  {\endlist}
  
\usepackage{xspace}

\usepackage{hyperref} 
\usepackage{tabularx} 

\begin{document}

\title[How Problematic Writer-AI Interactions (Rather than Problematic AI) Hinder Writers' Idea Generation]{How Problematic Writer-AI Interactions\\(Rather than Problematic AI) Hinder Writers' Idea Generation}


\settopmatter{authorsperrow=3}

\author{Khonzoda Umarova}
\authornote{Both first authors contributed equally to this research.}
\email{ku47@cornell.edu}
\orcid{0009-0003-7345-8741}
\affiliation{%
  \institution{Cornell University}
  \city{Ithaca}
  \state{New York}
  \country{USA}
}

\author{Talia Wise}
\authornotemark[1]
\email{tw294@cornell.edu}
\orcid{0000-0002-5031-0598}
\affiliation{%
  \institution{Cornell University}
  \city{Ithaca}
  \state{New York}
  \country{USA}
}

\author{Zhuoer Lyu}
\email{zl899@cornell.edu}
\orcid{0009-0004-9402-830X}
\affiliation{%
  \institution{Cornell University}
  \city{Ithaca}
  \state{New York}
  \country{USA}
}

\author{Mina Lee}
\email{mnlee@cs.uchicago.edu}
\orcid{0000-0002-0428-4720}
\affiliation{
  \institution{University of Chicago}
  \city{Chicago}
  \state{Illinois}
  \country{USA}
}

\author{Qian Yang}
\email{qianyang@cornell.edu}
\orcid{0000-0002-3548-2535}
\affiliation{%
  \institution{Cornell University}
  \city{Ithaca}
  \state{New York}
  \country{USA}
}


\begin{abstract}

Writing about a subject enriches writers' understanding of that subject.
This cognitive benefit of writing---known as \textit{constructive learning}---is essential to how students learn in various disciplines.
However, does this benefit persist when students write with generative AI writing assistants?
Prior research suggests the answer varies based on the type of AI, e.g., auto-complete systems tend to hinder ideation, while assistants that pose Socratic questions facilitate it.
This paper adds an additional perspective.
Through a case study, we demonstrate that the impact of genAI on students' idea development depends not only on the AI, but also on the students and, crucially, their interactions in-between.
Students who proactively explored ideas gained new ideas from writing, regardless of whether they used auto-complete or Socratic AI assistants. Those who engaged in prolonged, mindless copyediting developed few ideas even with a Socratic AI.
These findings suggest opportunities in designing AI writing assistants, not merely by creating more thought-provoking AI, but also by fostering more thought-provoking writer-AI interactions.

\end{abstract}

\maketitle

\section{Background} 

\begin{figure}[hb]
    \centering
    \vspace{-0.2cm}
    \includegraphics[width=0.36\paperwidth]{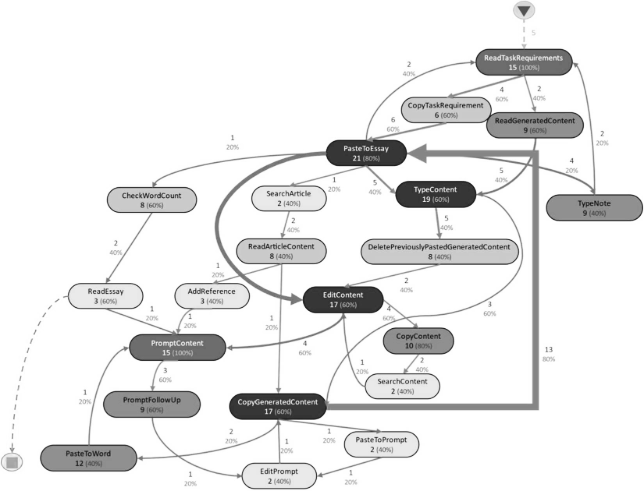}
    \caption{A visualization of one writer's thought process when writing an academic article~\cite{nguyen2024human}. Contrary to what theoretical models might suggest~\cite{flower_Hayes1981cognitive}, in practice, writers switch among various cognitive and metacognitive tasks ``\textit{in a seemingly haphazard fashion}” during writing~\cite{quinlan2012coordinating,wang2025scholawritedataset}, making it hard to analyze AI's impact on their thought processes.}

    \label{fig:MessyFlow}
    \Description{A visualization of one writer's thought process when writing an academic article. Writers shift among various cognitive and metacognitive tasks ``in a seemingly haphazard fashion” while writing, making it difficult to when and how AI suggestions have impacted their thought processes.}
\end{figure}

\begin{figure*}[hb]
    \includegraphics[width=0.6\textwidth]{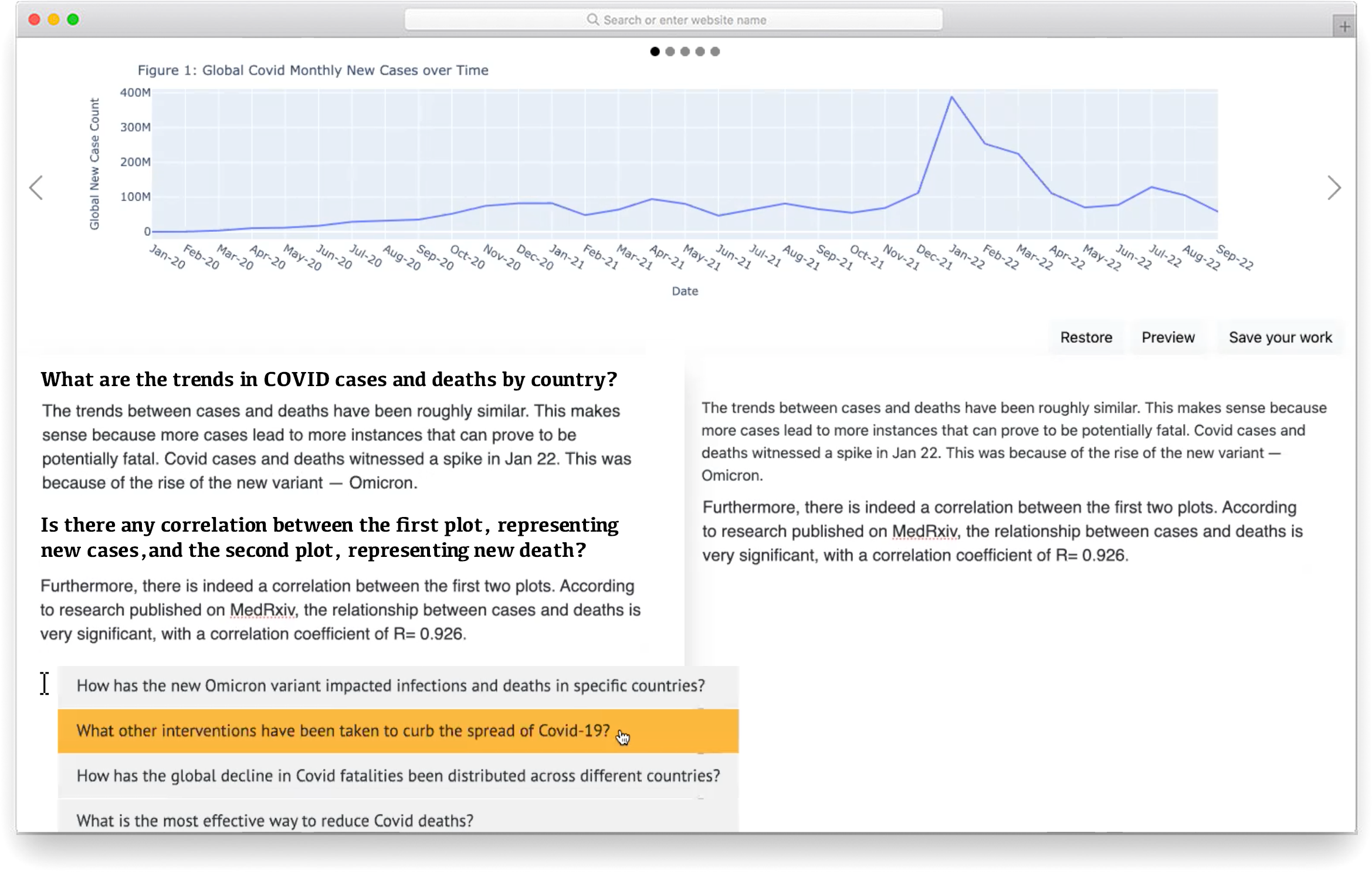}
    \caption{Screenshot of the text editor~\cite{Yang_coauthor_CHI22} and the Socratic AI writing assistant in use. Accepted AI suggestions/questions were displayed as an outline in the writing panel (left) and were invisible in the review panel (right).}
    \label{fig:UI}
\end{figure*}

The thought processes of writing about a subject help writers develop new ideas, a new understanding of that subject~\cite{emig2020writing,wittgenstein1982personal,emig1977writing_as_learning,langer1987writing_shapes_thinking,Klein2019_writing_as_learning}.
As simple as it might seem, this cognitive benefit of writing, known as \textit{constructive learning}, has been fundamental to U.S. educational practices for over fifty years~\cite{tynjala1998writing}.
It is why writing exercises are widely used to facilitate student learning not just in language-learning classes but across nearly all disciplines~\cite{bazerman2005writing_across_curriculum}. 

However, do students still gain new ideas, new understanding about their subject when they write with generative AI (genAI) writing assistants? 
This is an important question, as its answers would provide crucial evidence for educators and policymakers in deciding whether and how to embrace students' use of genAI writing assistants in constructive learning contexts~\cite{ParkCHI24_TeacherFear,HanCHI24Teachers}.
They also serve as a useful benchmark for designers who create novel AI writing assistants that enhance constructive learning.

Recent research has begun exploring this question.
This body of work found that, for example, writing with auto-complete systems tends to hamper idea generation, making writers less likely to gain original ideas about their subject through writing~\cite{anderson2024homogenization,Arnold_IUI20_preditable_text_predictable_writing,Poddar_CHI24EA,Li_CHI24_genAIWA,chakrabarty2023art_Arxiv,MauriceCHI23_OpinionatedGPT_bias_authors}.
Meanwhile, systems that pose Socratic questions can facilitate writers' idea generation~\cite{song_CHI2025_explore_self,SocraticAI_Orthopedic,chukhlomin2024socratic,favero2024Socratic_Chatbot_ECAIworkshop,lara2020_SocraticAI_moral_ethics}.

This paper adds to this line of research. We investigate not \textit{whether} writing with genAI writing assistants affects writers' idea generation, but \textit{when} and \textit{how} these impacts occur during the writing process.
We hope this more nuanced, mechanistic understanding can further inform educators' and policymakers' decisions regarding genAI adoption in education, and guide the design of novel genAI writing assistants that enhance constructive learning.

\section{Method}

\begin{figure*}[h!]
    \includegraphics[width=\textwidth]{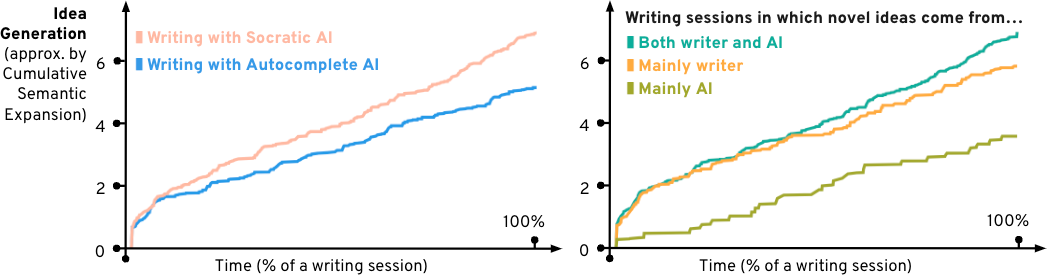} 
    \caption{The paces of semantic expansion in students' writing (a proxy for their idea generation) over the course of various writing sessions. While writing with the Socratic AI assistant resulted in greater semantic expansion compared to using auto-complete (left), the difference is much less substantial than the effect of the writers' active thinking and ideation (right).} 
    \label{fig:CogOutcomeAnalysis}
    \Description{The paces of semantic expansion in students' writing (a proxy for their idea generation) over the course of various writing sessions. While writing with the Socratic AI assistant resulted in greater semantic expansion compared to using auto-complete (left), the difference is much less substantial than the effect of the writers' active thinking and ideation (right).}
    \vspace{0.3cm}
\end{figure*}

To understand when and how genAI writing assistants might affect students' idea generation and constructive learning through writing, we undertook a four-step research process:
\begin{enumerate}[leftmargin=*]
    \item First, we chose a writing task known to engage students in idea generation and constructive learning;
    \item Next, we created two genAI writing assistants as probes: One designed to further enhance writers' idea generation and the other to hamper it;
    \item Then, we invited students to write articles using each of the AI assistants, while logging their writing behaviors and interviewing them about their thought processes;
\item Finally, we compared students' writing/thinking processes when enhanced versus hampered by AI assistants, thereby identifying AI's impacts.
\end{enumerate}

This innovative approach addresses a key challenge in evaluating AI's impact on writers' idea generation: the rich and transient nature of writers' thought processes. Neither logs nor interviews can perfectly capture these processes (Figure~\ref{fig:MessyFlow}), and no analysis can perfectly assess or quantify the ``\textit{ideas}'' that emerged from them.
Our approach addresses this challenge in two ways.
First,  we teased out AI's impact on writers' thinking/writing processes by comparing instances when AI either enhanced or hindered these processes.
Second, we made these comparisons by combining two imperfect but complementary data sources: (1) Writers' recollections of their thought processes, which offer rich insights but are highly subjective; and (2) writing process logs, which are less subjective, but require our interpretation to infer or quantify the emergence of writers' ideas.
We generated and confirmed the findings of this paper across both data sources to ensure rigor.

\paragraph{Writing Tasks.~}
We started by selecting a writing task widely known for engaging students in idea generation and constructive learning: \textit{Writing reflective essays based on a provided set of information}~\cite{trimbur2000composition}.
This task requires students to generate new ideas beyond the information provided, for example, by connecting it with prior knowledge to form a narrative or by constructing and defending a position. 

In this study, we invited students to write reflective essays on topics like climate change and gun control based on a provided set of information.
The instruction goes: ``\textit{Your goal is not to cover every piece of information provided, but to make your article as interesting and insightful as possible.}''~
Notably, the information provided to students in this study consisted solely of numerical data, presented as visualizations (Figure~\ref{fig:UI}). This allowed us to clearly distinguish between the information provided to the students and the genAI models (numbers) and the new ideas they generated (prose) when later analyzing the students' idea generation processes.
Appendix~\ref{appendix_writing_task} includes the full instruction.

\paragraph{Technology Probes.~}
We developed two generative AI ``\textit{assistants}" as probes: One is a GPT-3-based Socratic AI assistant. It is known to enhance writers' idea generation and learning~\cite{song_CHI2025_explore_self,SocraticAI_Orthopedic,chukhlomin2024socratic,favero2024Socratic_Chatbot_ECAIworkshop,lara2020_SocraticAI_moral_ethics}. 
The other is a GPT-3-based auto-complete system, known to hamper ideation generation~\cite{anderson2024homogenization,Arnold_IUI20_preditable_text_predictable_writing,Poddar_CHI24EA,Li_CHI24_genAIWA,chakrabarty2023art_Arxiv,MauriceCHI23_OpinionatedGPT_bias_authors}.
Upon request, the Socratic assistant generates four Socratic questions based on the data provided to students and the texts they have written so far~\cite{ho2023thinking,paul2019thinker,clark2015socratic}.
In contrast, the auto-complete system predicts four options for the next sentences, also based on the data provided and their writing so far. Appendix~\ref{appendix_prompt} details both systems' prompt designs. 

Both AI assistants presented their outputs through the interactions of a typical auto-complete system: When a student presses the tab key, the text editor displays four suggestions in a drop-down menu, allowing the student to accept one suggestion or dismiss them all (Figure~\ref{fig:UI}.)

\paragraph{Participants.~}
We recruited $29$ college students for this IRB-approved study.
We invited each participant to write two articles. For each article, they used a different AI assistant and focused on a different topic from a set of three: climate change, gun violence, and the COVID-19 pandemic.
The order of AI assistants and topics was randomly assigned.
Each writing session lasted 30 to 60 minutes and ended when the participant reached a natural stopping point. 

\paragraph{Data and Analyses.}
We collected both qualitative and quantitative data on the participants' writing/thinking processes. The qualitative data included the participants' recollections of their thought processes\footnote{All writing sessions were conducted over Zoom. During a session, a researcher screen-recorded a participant's writing process, turned off the camera, and stayed silent. Afterward, they conducted a retrospective-think-aloud interview with the participant~\cite{Yang_sketchingNLP_CHI19}, by replaying the screen recording and inviting participants to describe their thought processes at each moment.} as well as the texts they wrote. The quantitative data include keystroke-level logs of their writing processes and interactions with AI.

We confirmed all findings across both data sources.
Specifically, we manually identified new ideas about the subject matter as they appeared in participants' think-aloud scripts and written texts. We then hypothesized the influence of AI suggestions on these ideas and finally confirmed these hypotheses through quantitative analysis of the writing process logs.
For example, participant interviews suggested that prolonged copyediting might have slowed down their idea generation speed. To confirm this, we plotted the pace of semantic expansion\footnote{Appendix~\ref{appendix_semantic_distance} details how we computed the pace of semantic expansion.} in their written texts~(Figure~\ref{fig:CogOutcomeAnalysis}.) 

\section{Findings}

\begin{figure*}[hb]
    \includegraphics[width=\textwidth]{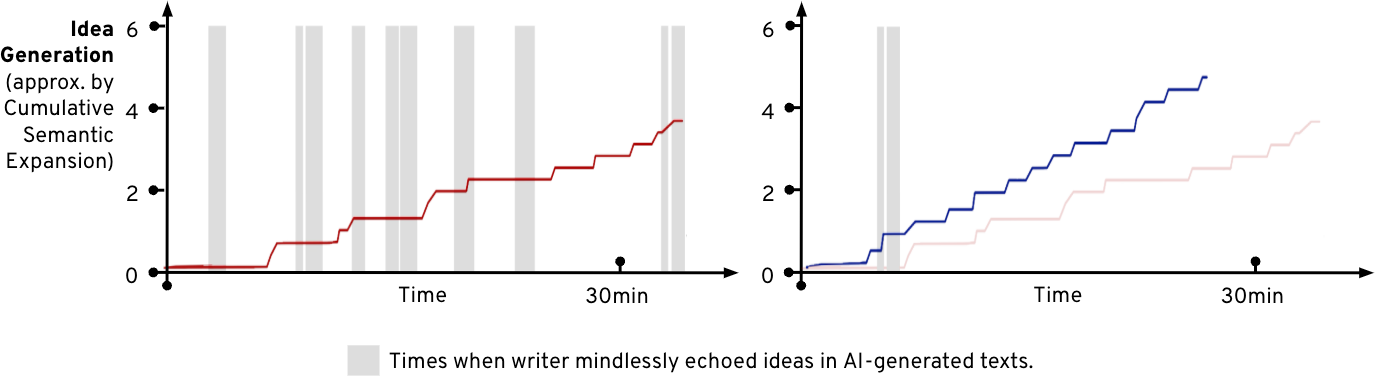} 
    \caption{When students mindlessly echoed AI suggestions, their idea generation decreased. The figure illustrates this by showing the paces of semantic expansion in a student's writing—a proxy for their idea generation—when they wrote with a typical auto-complete system (left) versus a modified auto-complete system that detects mindless echoing in real time and stops it by halting auto-complete suggestions for 2 minutes (right). When using the modified system, the student generated similar amount of new ideas in much shorter time. The gray areas in both figures represent periods of ``\textit{mindless echoing}". }
    \label{fig:MindlessEcho}
    \Description{TO COME.}
\end{figure*}

\subsection{Tools for Thought? It Takes Two to Tango.}

Our first finding is that our assumption based on prior research was wrong: Writing with auto-complete systems did not necessarily hinder participants' idea generation, nor did Socratic AI assistants necessarily facilitate it.
Instead, the impact of generative AI on idea generation depended on both the AI and the participants and, crucially, their interactions in between.

Figure~\ref{fig:CogOutcomeAnalysis}~illustrates this finding well. While writing with the Socratic AI assistant indeed resulted in more idea generation compared to using auto-complete systems (left figure), the difference is much less substantial than the effect of the participants' active engagement in the ideation process (right).
Overall, the writing sessions that generated the most new ideas were ``\textit{human-AI co-ideation sessions}'', writing processes where both the participant and the AI assistant contributed new ideas and built upon each other's ideas (right, green line).
Generating a nearly similar amount of new ideas is ``\textit{human-led sessions}'' where the participant wrote somewhat independently and generated new ideas without incorporating AI suggestions (right, orange line).
``\textit{AI-led sessions}'' where most new ideas came from AI suggestions produced the least amount of new ideas (right, yellow line). 

Participants who actively engaged in ideation developed new ideas through writing, regardless of whether they used the auto-complete or Socratic AI assistant.
These participants treated AI suggestions as ``\textit{additional information for brainstorming}"; therefore, even the mundane and occasionally off-topic AI suggestions could ``\textit{broadening [their] views on the problem.}"
P9's behaviors exemplified how this effect could occur. When P9 wrote about climate change, the auto-complete system offered an awkward next sentence suggestion: ``\textit{Cities should strive to create efficient road networks that present clear navigation[...].}" Interestingly, P9 did not accept or reject this suggestion straightaway. Instead, P9 asked themselves: What is this sentence trying to tell me? Is it encouraging me to think about how cities should create efficient road networks?
Following this line of questioning, P9 researched the relationship between city planning and climate change, expanding their writing in this direction, all prompted by an arguably erroneous next-sentence suggestion.

Participants who did not actively engage in ideation generated few new ideas, even with the help of a Socratic AI assistant. Out of the 29 participants, two exhibited this behavior.
They requested many Socratic questions at once as a way of brainstorming, sorted these questions by topic, and then edited them into prose.
Essentially, they manually converted the Socratic AI assistant's questions back into next-sentence suggestions, bypassing its benefits for idea generation. 




\subsection{Key Interactions in\\Writer-AI Collaborative Ideation}

We have demonstrated that an AI assistant's impact on students' idea generation depends not only on the AI itself but also on the \textit{interactions} between the student and the AI.
This finding has major implications.
For instance, educators and policymakers should consider not only choosing the AI assistants that most effectively stimulate learning, but also training students to engage with AI in a cognitively active manner.
Additionally, designers of new genAI writing assistants could design to foster more stimulating writer-AI interactions, rather than focusing solely on generating more thought-provoking AI outputs.

To jump-start such design efforts, below we identify three interactions pivotal to successful writer-AI collaborative ideation in our study. We encourage future research to further verify this crucial role across additional writing sessions and diverse contexts, as well as to identify any additional interactions. If these interactions are confirmed as pivotal, designers, educators, and policymakers can potentially use their occurrence as indicators of whether AI writing assistant designs enhance or compromise cognitive engagement.

\paragraph{Writer Mindlessly Echoing AI Outputs.~}
A clear indicator of student-AI collaborative ideation breakdown was ``\textit{mindless echoing}'', where a student's writing merely echoed the ideas from the AI assistant's outputs for a prolonged period of time.
Despite the student's continued production of new texts, the creation of new ideas has stopped.
We can identify such interactions from the logs of our AI-enhanced text editors~\cite{Yang_coauthor_CHI22} by detecting consecutive text insertion and/or deletion events that generated a large amount of text, without resulting in significant semantic expansion\footnote{\label{footnote_threshold}The thresholds for what is considered a "large" amount or a "significant" semantic expansion vary based on the writing topic, the length of the writing task, and the researchers' needs.}.

The occurrence of mindless echoing coincided with diminished idea generation in our participants' writing sessions.
This was the most evident when AI assistants generated nonsensical or irrelevant suggestions. Participants who were cognitively engaged easily recognized and dismissed these errors: ``\textit{No, actually I don't agree with that...This is exactly the opposite (of what I wrote earlier). So I canceled this.}'' (P13)
Some even could extract novel ideas from these errors (P9.)
However, those who mindlessly echoed the AI assistant (P4, P15) continued to build on the erroneous suggestions, often all the way until they could no longer follow what they were writing about.
One participant, having composed an entire article by merely sorting and echoing AI suggestions, finally read their own writing during the post-writing interviews. They laughed out loud: ``\textit{This doesn't make any sense!}''
---In sum, mindless echoing lets a single AI error escalate into a labyrinth of confusion, resulting in nonsensical articles.

\paragraph{Premature and Prolonged Copyediting.~}
Premature and prolonged copyediting also coincided with breakdowns in writer-AI collaborative ideation. In these scenarios, a participant spent excessive time making minor edits to existing text early in the writing process.
Despite a greater need to explore new ideas, they focused on fine-tuning the expressions and word choices of the few already-written ones.
We could identify such interactions from system logs by detecting long sequences of consecutive text insertion and/or deletion events that resulted in neither significant textual changes nor significant semantic expansion\footref{footnote_threshold}.

Premature and prolonged copyediting proved particularly problematic in AI-assisted writing. Early in the writing process, AI assistants could generate only bland and generic outputs, as they were prompted with nothing other than the writing task description and a dataset. 
By focusing on copyediting these anodyne texts, some participants wrote entire articles without ever forming a clear opinion or stance.  

\paragraph{Writer-Initiated Topic Shift.~}
Our study suggests that whether participants initiated topic shifts and how their interactions with the AI assistant unfolded afterward were crucial for successful writer-AI collaboration.
A lack of writer-initiated topic shifts---whether it was due to the participant merely echoing or copyediting the AI's suggestions---often indicated problems in idea generation.
Conversely, when a participant initiated an idea, stance, or position, it sometimes prompted the AI assistant to generate a more pointed suggestion that the participant then built upon. Through such interactions, writer-initiated topic shifts triggered a positive, self-reinforcing loop in writer-AI collaborative idea generation, resulting in some of the most productive writing sessions in our study.
\begin{quote}
    “\textit{I had a thought in the direction I wanted to go, [...] I was trying to get the AI to guide that thought.}" (P7)
    
    “\textit{I knew it wasn't gonna take me somewhere new unless I started typing [and] bringing up a new subject}. [...] So I baited the AI.” (P11)
\end{quote}

We identified moments of writer-initiated topic shift from system logs by detecting consecutive text insertion and/or deletion events that occurred after the start of a new sentence or paragraph (which we use as a proxy for topic shift) and led to minimal textual changes but substantial semantic expansion\footref{footnote_threshold}.

\section{Conclusion}

Through a case study, this paper demonstrated that the impact of genAI on students' idea generation depends not only on the AI but also on the students and, crucially, their interactions.
Further, in our study, we identified three interactions that crucially shaped AI's impact on the writers' idea generation. We offered not just the qualitative descriptions of these interactions but also ways to identify them from the logs of AI-enhanced text editors.

We hope this work could compel more researchers, designers, educators, and policymakers to see genAI's impact on students' writing-to-learn processes through the lens of interactions. We invite more researchers and designers to join us in exploring how AI writing assistants could foster more thought-provoking writer-AI interactions.

\begin{acks}
We thank Swati Mishra and Zhongqian Li for their help with the earlier iterations of this study.

Qian Yang's effort was supported in part by the AI2050 Early Career Fellowship program at Schmidt Sciences.
This material is also based upon work supported by the National Science Foundation under Grant No. 2313078. Any opinions, findings, and conclusions or recommendations expressed in this material are those of the author(s) and do not necessarily reflect the views of the National Science Foundation.
\end{acks}

\bibliographystyle{ACM-Reference-Format}
\bibliography{ref/misc,ref/PIprior,ref/theraputic_writing,ref/writing,ref/constructive_learning_writing}

\appendix

\section{Writing Tasks for Participants}\label{appendix_writing_task}
The instructions provided to participants were as follows:
\begin{quote}
    You are tasked with writing an article about the issue of {gun violence in the U.S.} for {name of a local newspaper}. A new dataset on this topic has just been released, presented in the form of these five visualizations. 
    You want your reporting to stand out, among the many people who got their hands on the dataset and are writing for other newspapers. Consider, for example, whether you can find a new angle that others might not notice, or write more deeply about an angle. To be clear, your goal is not to cover every piece of information in the dataset, but to make your article as interesting and insightful as possible. You should draw on the AI-generated suggestions for your writing, and you can also use Google if you need any extra information.
\end{quote}

\section{Prompts for Writing Assistants}\label{appendix_prompt}

First, we prompted both writing assistants with a description of the data given to participants.
For example, the following prompt describes the gun violence data.

\texttt{Analyze the following data: We have 5 visualizations generated from the data.The first plot is a time series of gun violence incidents over time from Feb 2013 to March 2018. The y-axis is the number of incidents, and the x-axis is the time period. The second plot is a bar chart of gun violence incident counts per state. The y-axis represents the number of incidents, and the x-axis has the states sorted high-to-low in incident counts. The third plot is a stacked bar chart of injured and killed people by each state. The two variables are the number of people injured and the number of people killed. The y-axis is the victim count, and the x-axis has the states sorted in alphabetical order. The fourth plot is a stacked bar chart of victim counts by gender by each state. The two variables are male and female victim counts. The y-axis is the victim count, and the x-axis has the states sorted in alphabetical order. The fifth plot is a stacked bar chart of children and teen victim counts by each state. The two variables are children and teen victim counts. The y-axis is the victim count, and the x-axis has the states sorted in alphabetical order.}

\paragraph{Prompt for Socratic AI Assistant.~}
The Socratic AI assistant also includes the following prompt:

\texttt{{The last 10 sentences preceding the cursor} Based on the text above, ask four Socratic questions on what has not yet been addressed in the writing. Socratic questions lead to exploring complex ideas, uncovering assumptions,  and analyzing concepts. Examples of Socratic questions include: 'What are the alternative explanations for the trend of increasing gun violence incident counts', 'What are the implications of discrepancy in energy consumption profiles?', or 'What evidence supports the claim of weather conditions contributing to road safety?'. Please ask four questions in the following format: 1. [QUESTION 1] 2. [QUESTION 2] 3. [QUESTION 3] 4. [QUESTION 4]}

We retrospectively confirmed that the questions generated by the Socratic AI aligned with the instructions in the prompt.
\begin{itemize}[leftmargin=*]
    \item The questions were indeed Socratic. While only 163 (28.55\%) of them followed the Socratic question template patterns \cite{paul2019thinker,socraticQgen_ACL23} verbatim, all were clearly reminiscent of the templates.

    \item The questions were relevant to the proceeding texts. The semantic similarity scores between each question and the proceeding texts are 0.75 (measured by word2vec) and 0.79 (BERT embeddings.)
    \item The questions identified gaps in the students' writing so far. 
    We confirmed this based on the fact that the questions generated by the Socratic AI assistant were semantically very different from the questions that the writers' writing so far can answer (p-value<<0.01).
    Specifically, we compare questions (Q) generated by our system to questions ($Q'$) generated by a state-of-the-art question generation system intended for developing question answering models~\cite{du2017learning}. For each writing context (c), we generate both Q and $Q'$ questions. We calculate their semantic similarity $sim(q_c, q'_c)$, where $q_c\in Q$ and $q'_c\in Q'$ for context c. We observe that for the same context c, questions in Q are different from questions in Q` (p-value$<<0.01$).
\end{itemize}

\paragraph{Prompt for the auto-complete AI Assistant.~}
The auto-complete AI assistant also includes the following prompt: 

\texttt{{The last 10 sentences preceding the cursor} Based on this context, suggest next sentences in the following format: 1. [SENTENCE1] 2. [SENTENCE2] 3. [SENTENCE] 4. [SENTENCE]}

\section{Quantifying Idea Generation Processes}\label{appendix_semantic_distance}
We captured a ``screenshot'' of the article being written at various points during the writing process (ex. when the writer finishes inserting text and moves the cursor or decides to request AI suggestions). We then calculated the semantic distance between each pair of consecutive snapshots.
Let snapshot $S_i$ be the text content within the editor at point in time $i$. Let $n_i$ and $n_{i-1}$ be the number of sentences in snapshots $S_i$, $S_{i-1}$ respectively.
From here, we define:
\[semanticExpansion(i)=1-\frac{sim(S_i, S_{i-1})}{|n_i-n_{i-1}|+1},\]
To account for difference in the amount of text, we normalize the similarity score by the number of sentences changed between the two snapshots. And finally, we subtract the result from 1 so that the score quantifies difference between writing snapshots.
In our analyses, we used the similarity score based on word2vec embeddings~\cite{mikolov2013efficient}. 

\end{document}